\documentclass[twocolumn,amsmath,amssymb,aps,superscriptaddress,nofootinbib]{revtex4-2}

\usepackage[english]{babel}
\usepackage[utf8]{inputenc}
\usepackage{xr-hyper}

\usepackage[colorlinks]{hyperref}
\usepackage[T1]{fontenc}
\usepackage[dvips]{graphicx}
\usepackage{amsmath}
\usepackage{amsfonts}
\usepackage{color}
\usepackage{ulem}
\usepackage[caption=false, justification=justified]{subfig}
\usepackage{url}
\usepackage{ulem}
\usepackage[top=2cm, bottom=2cm, left=2cm, right=2cm]{geometry}

\makeatletter
\newcommand*{\addFileDependency}[1]{
  \typeout{(#1)}
  \@addtofilelist{#1}
  \IfFileExists{#1}{}{\typeout{No file #1.}}
}
\makeatother

\newcommand*{\myexternaldocument}[1]{%
    \externaldocument{#1}%
    \addFileDependency{#1.tex}%
    \addFileDependency{#1.aux}%
}

\myexternaldocument{supplementary_information_PRL}

\hypersetup{filecolor=blue}
\hypersetup{citecolor=blue}
\hypersetup{urlcolor=blue} 

\hypersetup{linkcolor=blue}

\newcommand{\sech}{\operatorname{sech}}

\begin{document}

\title{Large interfacial Rashba interaction and giant spin-orbit torques in atomically thin metallic heterostructures}

\date{\today}

\author{S. Krishnia}
\author{Y. Sassi}
\author{F. Ajejas}
\author{N. Reyren}
\author{S. Collin}
\author{A. Fert}
\author{J.-M. George}
\author{V. Cros}
\email{vincent.cros@cnrs-thales.fr}
\author{H. Jaffrès}
\email{henri.jaffres@cnrs-thales.fr}
\affiliation{Unité Mixte de Physique, CNRS, Thales, Université Paris-Saclay, 91767, Palaiseau, France}

\begin{abstract}\textbf{
The ability of spin-orbit interactions to convert charge current into spin current, most often in the bulk of heavy metal thin films, has been the hallmark of spintronics in the last decade. In this study, we demonstrate how the insertion of light metal element interface profoundly affects both the nature of spin-orbit torque and its efficiency in terms of damping-like ($H_{\text{DL}}$) and field-like ($H_{\text{FL}}$) effective fields in ultrathin Co ferromagnet. Indeed, we measure unexpectedly large $H_{\text{FL}}$/$H_{\text{DL}}$ ratio ($\sim$2.5) upon inserting a 1.4 nm thin Al layer in Pt|Co|Al|Pt as compared to a similar stacking including Cu instead of Al. From our modelling, these results strongly evidence the presence of large Rashba interaction at Co|Al interface producing a giant $H_{\text{FL}}$, which was not expected from a metallic interface. The occurrence of such enhanced torques from an interfacial origin is further validated by demonstrating current-induced magnetization reversal showing a significant decrease of the critical current for switching. } 
\end{abstract}

\maketitle


In spinorbitronics~\cite{Soumyanarayanan2016}, the spin-orbit coupling (SOC) at interfaces is known and exploited to generate strong interfacial perpendicular anisotropy (PMA) or antisymmetric exchange interactions. Moreover, SOC also generates efficient spin-current \textit{via} the spin Hall effect (SHE), that occurs in the bulk of heavy metal thin films, used to switch the magnetization~\cite{manchon2015,manchon2019} or to drive chiral spin textures as \textit{e.g.} domain walls~\cite{miron2010,miron2011} and skyrmions~\cite{Fert2017} in heavy-metal/ferromagnet stacks. In the case of atomically thin layers with broken inversion symmetry, the interplay of low dimensionality and SOC may also lead to Rashba spin splitting ~\cite{rashba1983,miron2010} able to modify the electronic ground state at interfaces thus affecting the spin transport through the Rashba-Edelstein effect (REE)~\cite{edelstein1990}. These SOC-related effects, namely SHE~\cite{hirsh1999,sinova2015,manchon2019} and REE~\cite{edelstein1990,sinova2015,sanchez2013,manchon2015,manchon2019}, possess the property to convert a charge current $\mathcal{J}_c$ into either an out-of-equilibrium spin-current $\mathcal{J}_\sigma$ or a spin accumulation $\hat{\mathbf{\mu}}$ carrying angular momentum. Upon current injection, spin-accumulation and the spin-current can in turn transfer their angular momentum to the adjacent ferromagnetic layer by precessing around the exchange field. 

To conserve the total angular momentum, the local magnetization $\mathcal{M}=M\hat{\mathbf{m}}$ ($\Hat{\mathbf{m}}$ is the direction of the local magnetization and $M$ its norm) experiences the so-called spin-orbit torques (SOTs). 
Both $\mathcal{J}_c$ and $\hat{\mathbf{\mu}}$ may act in concert to exert net SOTs onto a magnetization vector $\hat{\mathbf{m}}$ with two different components: the damping-like torque (DLT), $\hat{\tau_{\text{DL}}} \propto \hat{\mathbf{m}} \times (\hat{\mathbf{\mu}} \times \hat{\mathbf{m}})$ and the field-like torque (FLT), $\hat{\tau_{\text{FL}}} \propto \hat{\mathbf{\mu}} \times \hat{\mathbf{m}}$ giving rise correspondingly to damping-like ($H_{\text{DL}}$) and field-like ($H_{\text{FL}}$) effective fields. Apart from identifying the actual source of spin-current, the exact nature and the relative amplitudes of the two SOT components strongly depend also on the relaxation length of transverse spin component, \textit{i.e.} spin decoherence length, which is predicted to be in the range of a few atomic lattice parameters~\cite{taniguchi2008,gosh2012}. Manipulating the magnetization direction \textit{via} SOTs and disentangling the exact origin of SOTs, and thus being eventually able to use them adequately, is still today a complex problem, notably in case of very thin films. 

In the pioneer experiments by Miron \textit{et al} in Pt|Co|AlO$_x$~\cite{miron2011}, the origin of the spin generation responsible for the magnetization switching was pointed out to be predominantly through FLT from REE at the Co|Oxide interface (larger than SHE-DLT from the Pt layer), suggesting the crucial role of a large interfacial electric field. Nevertheless, in all metallic heterostructures ~\cite{stiles2003,haney2013,amin2016,amin2016b,liu2012}, SHE is largely considered as the primary source of DLT, even though a large interfacial REE has been measured recently in Ir|ferromagnetic|Ta stacks that in addition contributes to the DLT~\cite{ralph2021}. Note that other physical mechanisms might also influence the relative amplitude (and thus the nature) of the two SOT components in the limit of atomically thin ferromagnetic layer i.e., spin-filtering or quantum confinement of charges leading to subsequent change of the electronic band structure~\cite{Qiu2015,emori2016,Amin2018,Wang2019,Qiu2015}, for example surface states of topological insulators (TIs)~\cite{Mahendra2018} and in 2D electron gases ~\cite{Noel2020}. 
 
Here, we have investigated the properties of SOTs in atomically thin metallic structures, namely Co layers ranging from 1.4~nm down to 0.4~nm, sandwiched between Pt and light element based (Al, Cu) overlayers. Using the second harmonic Hall technique \cite{krishnia2021,Garello2013}, we precisely determine the amplitude of different SOT contributions as a function of the thickness of the ferromagnetic film, the light-element overlayer as well as the heavy-element film. Compared to the best-reported results ~\cite{Garello2013}, we find an increase of the DLT by $\sim$30~\% together with a drastic rise the FLT by more than $\sim$190~\%. By combining these experimental results with our spin-dependent Boltzmann calculations, we explicitly demonstrate the existence of an enhanced SOT mainly due to a strong REE in asymmetric Pt|Co|Al|Pt stacks, as evidenced by a FLT over DLT ratio as large as 2.5 for 0.55~nm thick Co, the highest value ever reported in metallic systems.



\begin{figure}
         \centering
         \includegraphics[width=0.45\textwidth]{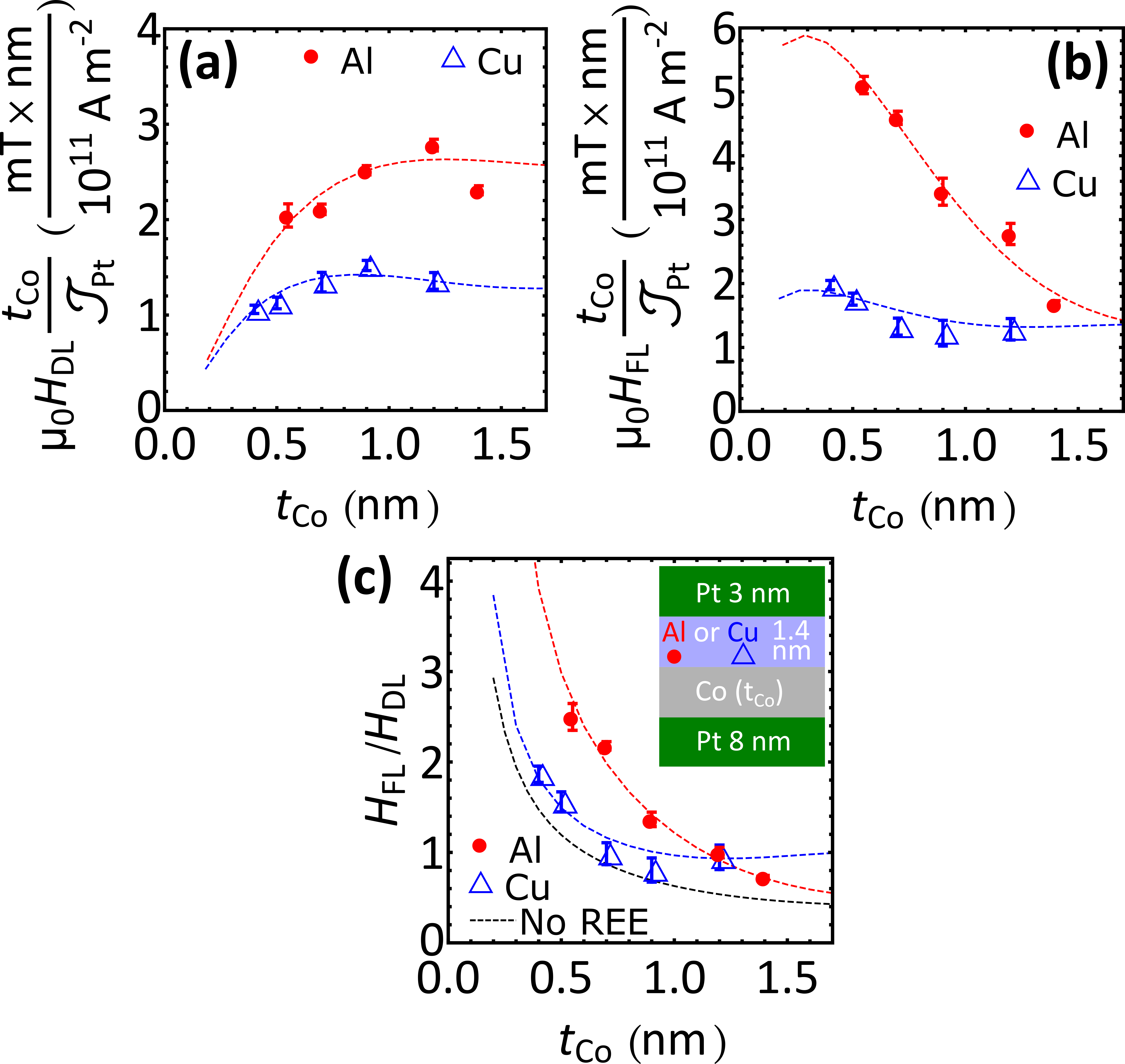}
\caption{a) DL field and (b) FL field multiplied by $t_{ Co}$ for $\mathcal{J}_{\text{Pt}}$ = $10^{11}$~A/m$^2$ in Pt(8)|Co(0.55-1.4)|Al(1.4)|Pt(3) and Pt(8)|Co(0.4-1.2)|Cu(1.4)|Pt(3). c) $t_{\rm Co}$ dependence of $\zeta=H_{\text{FL}}/H_{\text{DL}}$. The red and blue dashed lines are fits to experimental data points using our theoretical model and the black dashed line in (c) are calculated $\zeta$ values without any Rashba interface~(\textcolor{blue}{see SI-VII}).}
\label{Fig2}
\end{figure}


For this study, two series of multilayers have been grown having Al and Cu light element interface with Co: Pt(8)|Co($t_{\rm Co}$)|Al(1.4)|Pt(3) and Pt(8)|Co($t_{\rm Co}$)|Cu(1.4)|Pt(3) using magnetron sputtering. All the samples show perpendicular magnetic anisotropy. Starting from negligible torque in fully symmetric Pt(8)|Co(0.9)|Pt(8) stack (\textcolor{blue}{SI-IV}), the asymmetric \textit{'control'} sample Pt(8)|Co(0.9)|Pt(3) gives $H_{\text{DL}}$=0.80~$\pm$ 0.05~mT and $H_{\text{FL}}$ 0.58$\pm$0.15~mT for a current density $\mathcal{J}_{\text{Pt}}$ in Pt = $10^{11}$~A/m$^2$, and we will use it as a reference in the following. To obtain an integrated torque in the whole ferromagnetic layer of thickness $t_{\rm Co}$ and hence to be able to make accurate comparison for the sample series, we define a normalized quantity by multiplying the $H_{DL,FL}$ with $t_{\rm Co}$ for a $10^{11}$ A/m$^2$ current density in Pt. Note that the Cu series will be used later on as reference SHE sample, i.e. free of significant REE contribution.

Several features can be highlighted showing similarities and differences between the Al-based and Cu-based series. First, as shown in Fig.~\ref{Fig2}a, the integrated DL component $(H_{\text{DL}} \times t_{\rm Co})/ \mathcal{J}_{\text{Pt}}$ is about 65\% larger in Pt|Co|Al(1.4)|Pt samples if compared to Pt|Co|Cu(1.4)|Pt for almost the whole thickness range. At $t_{\rm Co}=0.9$~nm, the value of DLT is about 2.5~mT$\times$nm/($10^{11}$Am$^{-2}$) for Al and 1.5~mT$\times$nm/($10^{11}$~Am$^{-2}$) for Cu. Moreover, $(H_{\text{DL}} \times t_{\rm Co})/\mathcal{J}_{\text{Pt}}$ exhibits a similar qualitative maximum in the $1.0-1.2$~nm Co thickness window. Second, we find that, in both series, FLT exhibits an increase on reducing $t_{\rm Co}$ (Fig.~\ref{Fig2}b), resulting in a dominating FLT component at small $t_{\rm Co}$. Even if the SOT thickness dependence for Al(1.4) and Cu(1.4) series show similar trends, their magnitude strongly differs. We observe a substantial increase in the FLT for Al(1.4) series \textit{i.e.} from $1.7$~mT$\times$nm/($10^{11}$~Am$^{-2}$) for $t_{\rm Co}=1.4$~nm to $5.2$~mT$\times$nm/($10^{11}$~Am$^{-2}$) for $t_{\rm Co}=0.55$~nm, whereas the maximum value of FLT reaches only $2$~mT$\times$nm/($10^{11}$~Am$^{-2}$) for Cu. Such a large increase in atomically thin Co layer has never been reported before up to our knowledge. This striking result is manifested by the FLT to DLT amplitude ratio $\zeta=\frac{H_{\text{FL}}}{H_{\text{DL}}}$ found for Pt|Co|Al|Pt and considered as a new figure of merit (Fig.~\ref{Fig2}c). 

A rise in the FLT is expected \textit{i}) in the case of SHE, when the spin decoherence length is larger than the $t_{Co}$ as calculated and shown by the black dotted curve in Fig.~\ref{Fig2}c,  and \textit{ii}) due to an increased weight of the REE effects. In details, $\zeta=\frac{H_{\text{FL}}}{H_{\text{DL}}}$ saturates for $t_{\rm Co}\gtrsim1.2$~nm with a  magnitude $\zeta\approx~0.7$. This is indeed expected in the limit of a 'thick' ferromagnet~\cite{stiles2003} when only SHE is considered as for the Cu series where no Rashba is expected. Note that similar value of $\zeta$ were also obtained in our previous studies on 4~nm thick Co|Ni ferromagnetic multilayer~\cite{figuereido2021}. Reducing $t_{\rm Co}$ below 1.2~nm, we see in Fig.~\ref{Fig2}c that $\zeta$ increases sharply for Pt|Co|Al|Pt series and eventually reaches at 2.5 for $t_{\rm Co}\approx0.55$~nm, whereas remaining smaller than unity down to $t_{\rm Co}\approx0.7$~nm for Pt|Co|Cu|Pt. Such a large SOT enhancement by a factor larger than three indeed strongly challenges the SHE origin of SOT in samples involving Al. 


\begin{figure}
         \centering
         \includegraphics[width=.45\textwidth]{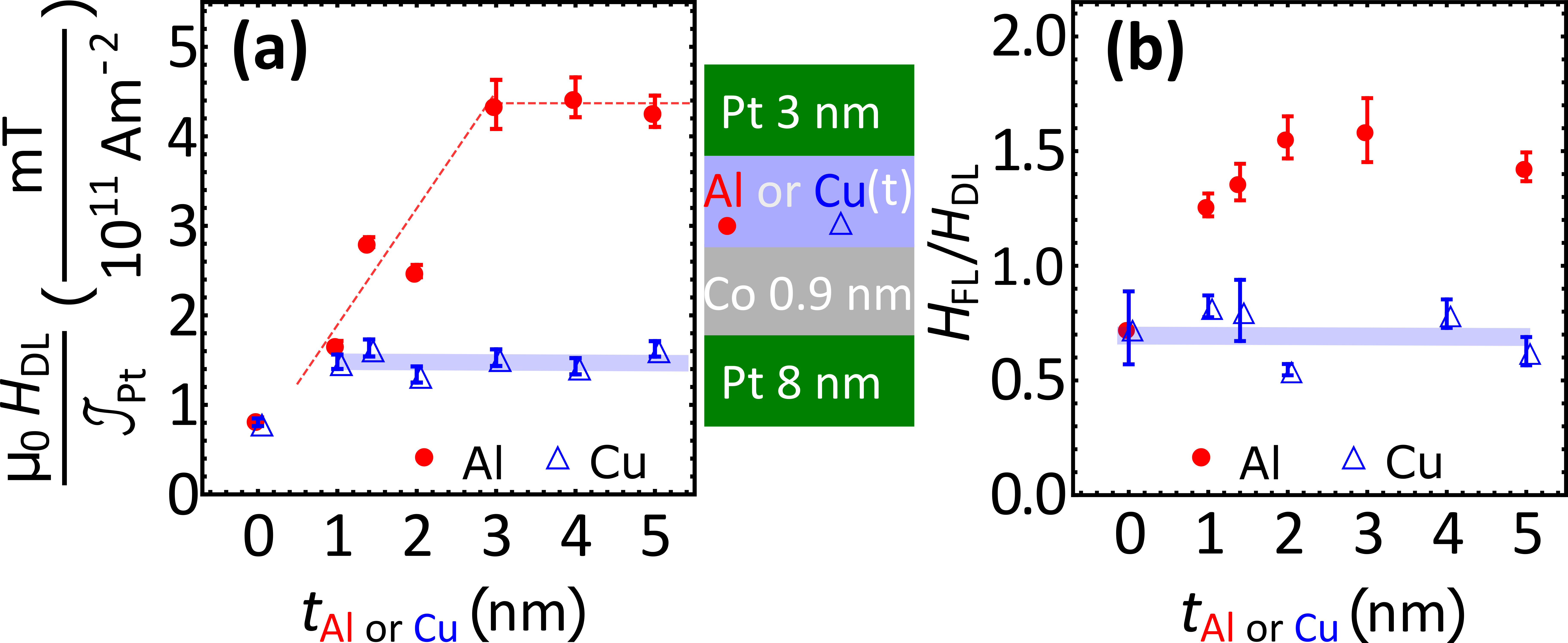}
\caption{a) $H_{\text{DL}}$ field  \textit{vs.} Al (red) and Cu (blue) layer thickness. The variation in $H_{\text{DL}}$ is fitted by the dashed red lines (\textcolor{blue}{see SI-VII}). We used are $l_{\text{sf}}^{\text{Pt}}$=1.5~nm, $\tilde{r}_s=2$, $\mathcal{L}~\theta_{\text{SHE}}^{\text{Pt(0)}}=0.09$ and $\mathcal{T}$ varying from $\mathcal{T}$=0.8 for $t_{\text{Al}}=0$~nm to $\mathcal{T}$=0 for $t_{\text{Al}}>3$~nm.  b) $\zeta=H_{\text{FL}}/H_{\text{DL}}$ \textit{vs.} Al (red) and Cu (blue) layer thickness. The thick and constant blue lines are guide to the eye to indicate that SOTs are independent of Cu thickness.}
\label{Fig3}
\end{figure}

In Fig.~\ref{Fig3}, we present the evolution of $H_{\text{DL}}/\mathcal{J}_{\text{Pt}}$ and $\zeta=\frac{H_{\text{FL}}}{H_{\text{DL}}}$ with the light element thickness \textit{i.e.} Al (red points) and Cu (blue points). We notice that the DLT exhibits an unexpected strong increase with $t_{\text{Al}}$ in the 1-3~nm window from $\sim1.5$~mT/($10^{11}$A/m$^2$) for $t_{\text{Al}}$ = 1~nm to a saturation value $\sim4.5$~mT/($10^{11}$A/m$^2$) for $t_{\text{Al}}\geq$ 3~nm. In parallel, the ratio $\zeta$ also increases much more with $t_{\text{Al}}$ (Fig.~\ref{Fig3}b) from $\zeta=0.75$ without Al to $\zeta=1.75$ for $t_{\text{Al}}\geq$ 3~nm, manifesting again a peculiar behavior. This trend is very different in the 'reference' Cu series with varying $t_{\text{Cu}}$. The $H_{\text{DL}}/\mathcal{J}_{\text{Pt}}\approx$~1.5~mT/($10^{11}$A/m$^2$) and $\zeta~\approx$~0.7 remain roughly constant \textit{vs.} $t_{\text{Cu}}$, as shown by blue open triangles in Fig.~\ref{Fig3}a and Fig.~\ref{Fig3}b. This corresponds to a rise of the FLT by at least a factor of 6 by inserting 2~nm Al on top of Co. 

Two major conclusions can be draught from these results: \textit{i}) a large electronic transmission across the top Co|Cu|Pt interfaces~\cite{jaffers2014,zhang2015} partially compensates the SHE torque from the bottom Pt, irrespective of $t_{\rm Cu}$ and ii) a progressive extinction of SHE action from the top Pt with increasing the thickness of 'Al barrier'\cite{footnote}. The effective spin Hall angle ($\theta_{\text{SHE}}^{\text{eff}}$) taking into the contributions from both the bottom and top Pt layers can be written:
\begin{equation}
   {\theta_{\text{SHE}}^{\text{eff}}}\approx{\theta_{\text{SHE}}^{\text{bulk Pt}}~[1-\mathcal{L}][~\mathcal{F}(8,\mathcal{T}_{\text{Co|Pt}})-\mathcal{F}(3,\mathcal{T}_{\text{Co|Al|Pt}})]},
\label{equ:efficiency_SHE}
\end{equation}
where $\theta_{\text{SHE}}^{\text{bulk Pt}}$ is the spin Hall angle of bulk Pt, $\mathcal{T}$ is the transmission coefficient, $\mathcal{L}$ is the spin memory loss~\cite{jaffers2014,berger2018,buhrman2019} at the interface. The function $\mathcal{F}$, defining the spin-back flow in Pt, is given by:
\begin{equation}
    \mathcal{F}\left(t_{\rm Pt},\mathcal{T}\right)\approx \frac{\mathcal{T}\tilde{r}_s^{\text{Pt}}\coth\left(\frac{t_{\text{Pt}}}{\lambda_{\text{sf}}^{\text{Pt}}}\right)\left[1- \sech\left(\frac{t_{\text{Pt}}}{\lambda_{\text{sf}}^{\text{Pt}}}\right)\right]}{1+\mathcal{T}\tilde{r}_s^{\text{Pt}}\coth\left(\frac{t_{\text{Pt}}}{\lambda_{\text{sf}}^{\text{Pt}}}\right)}
\label{equ:efficiency_SHE}
\end{equation}
$\lambda_{\text{sf}}^{\text{Pt}}$ is the spin-diffusion length (SDL) in Pt and $\tilde{r}_s^{\text{Pt}}$ is the spin-resistance or resistance to spin-flip (\textcolor{blue}{SI-VII}). From our fit shown in Fig.~\ref{Fig3}a, we extract $\lambda_{\text{sf}}^{\text{Pt}}\approx$ 1.5~nm in agreement with Ref.~\cite{zhang2015}, $\tilde{r}_s^{\text{Pt}}\approx$ 2, $\theta_{\text{SHE}}^{\text{Pt}}=(1-\mathcal{L})~\mathcal{T}~\theta_{\text{SHE}}^{\text{bulk Pt}}=0.09$ and $\theta_{\text{SHE}}^{\text{Bulk Pt}}$~$\approx$ 0.22 in agreement with the literature for Pt~\cite{zhang2015,dang2020} of resistivity $\rho=25 \mu\Omega$cm but smaller than 0.4~\cite{berger2018}.  We consider $\mathcal{T}=0.8$ and $\mathcal{L}\approx 0.5$. Similar values of $\theta_{\text{SHE}}^{\text{Bulk Pt}}$ were determined from our AHE~\cite{dang2020} measurements. Note that the SHE compensation from the top injection then approaches $\frac{2}{3}$ in our 'control' Pt(8)|Co(0.9)|Pt(3) structure. We can hence conclude that the large increase of $H_{\text{FL}}$ as well as $\zeta$ for Al cannot simply be explained owing to such bottom/top SHE compensation, suggesting the existence of a large REE for Co|Al.

\begin{figure}
         \centering
         \includegraphics[width=.45\textwidth]{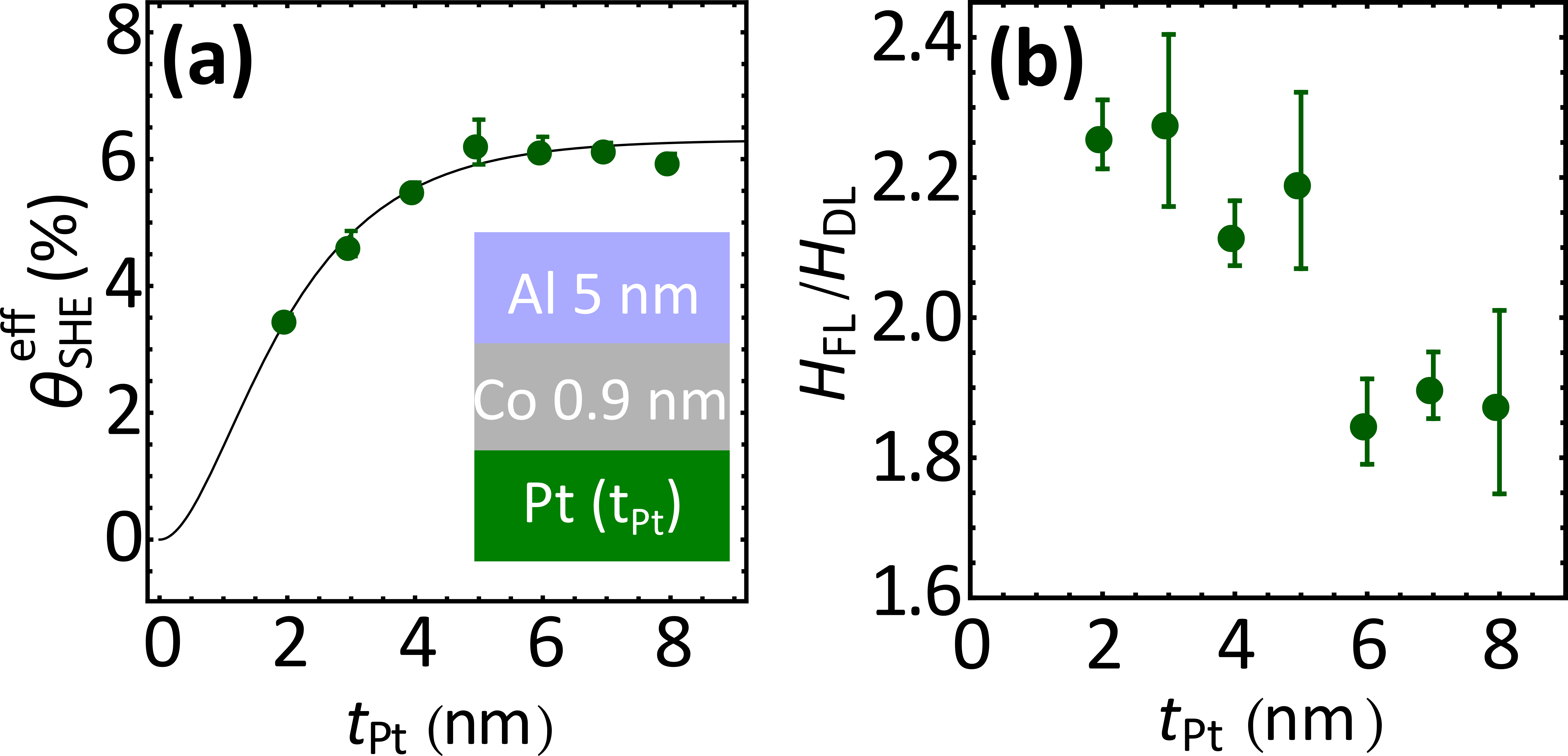}
\caption{a) Effective $\theta_{\text{SHE}}^{\text{eff}}$ \textit{vs.} $t_{\text{Pt}}$ (green points). The solid line is the result of the fit performed enable to extract $l_{\text{sf}}^{\text{Pt}}=1.5$~nm, bulk $\theta_{\text{SHE}}^{\text{eff}}=0.09$ of Pt and $\tilde{r}_s^{\text{Pt}}$=2 using Eq.~\ref{equ:efficiency_SHE}. b) The $\zeta$ ratio as a function of $t_{\text{Pt}}$.}
\label{Fig4}
\end{figure}

In order to discard the different contributions to the SOTs, our approach has been to control the amplitude of the SHE component that can be achieved through the decrease of the bottom Pt thickness in Pt($t_{\text{Pt}}$)|Co(0.9)|Al(5) series. We first quantify the SHE contribution from Pt by extracting $\theta_{\text{SHE}}$ from the DL field according to $\theta_{\text{SHE}}^{\text{eff}}=\frac{2e}{\hbar}\frac{\mu_0 H_{\text{DL}}}{\mathcal{J}_{\text{Pt}} M_\text{s} t_{\rm Co}}$ ($e$ is the electron charge, $M_\text{s}$ the saturation magnetization). In Fig.~\ref{Fig4}a, we show that $\theta_{\text{SHE}}^{\text{eff}}$ increases with $t_{\text{Pt}}$ according to the standard law $\propto \theta_{\text{SHE}}^{\text{Bulk Pt}}\times \mathcal{F}\left(t_{\text{Pt}}\right)$ (see. Eq.[\ref{equ:efficiency_SHE}]) before saturating for $t_{\text{Pt}}\geq5$~nm. From the fit, the spin-diffusion length (SDL) and $\theta_{\text{SHE}}^{\text{eff}}$ are $\simeq$~1.5~nm and $\simeq$~0.09 in agreement with data of Fig.~\ref{Fig3}. 
In Fig.~\ref{Fig4}b, we show the impact of REE from the Co|Al interfaces on the SOTs by restricting to $t_{\rm Co}=0.9$~nm. For this Co thickness, FLT dominates DLT for all Pt thickness. In addition, a remarkable increase of $\zeta$ is observed in the $t_{\text{Pt}}=$~2-5~nm range, \textit{i.e.} whereas SHE has not yet reached its maximum efficiency. This undoubtedly supports the role of a significant Co|Al Rashba interface, being superimposed to SHE. Our data also yield a clear quantitative understanding of how the Rashba effect at metallic interfaces may strongly affect $H_{\text{FL}}$ in chemically asymmetric Pt|Co|Al|Pt stacks as also described in Fig.~\ref{Fig2} and Fig.~\ref{Fig3}. 


\begin{figure}
         \centering
         \includegraphics[width=0.45\textwidth]{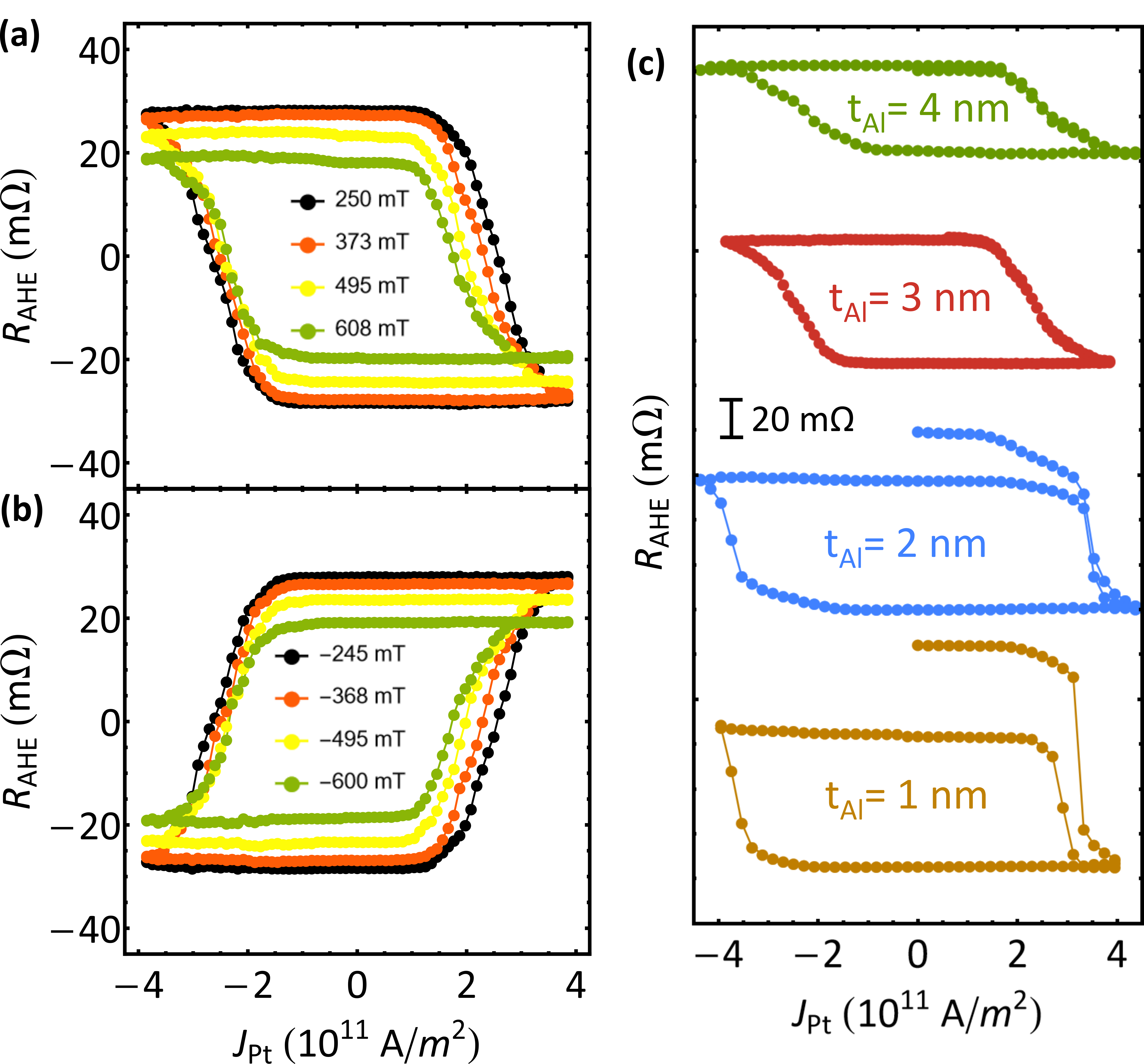}
\caption{Al thickness dependence of current-induced magnetization reversal: (a) Current-induced magnetization reversal in Pt(8)|Co(0.9)|Al(3)|Pt(3) for different (a) positive and (b) negative $H_x$. The inversion of R-I loops with field polarity confirms that SOT are at play. (c) R$_{AHE}$ as a function of pulsed current in Pt(8)|Co(0.9)|Al($t_{Al}$)|Pt(3) series of samples for $H_x=+250$~mT in each case. All the measurements were performed at 300~K. Pulse width was 100 $\mu$s and R$_{AHE}$ was measured after a 40 $\mu$s delay.}
\label{Fig5}
\end{figure}

To further demonstrate of the REE contribution to the SOTs from Co|Al interface, we perform SOT-driven magnetization switching experiments in series of Pt(8)|Co(0.9)|Al($t_{Al}$)|Pt(3) Hall cross-bars. As usually done, an external magnetic field $H_x$, either positive of negative, is applied along the current direction to break the mirror symmetry and $R_{AHE}$ is measured as a function of current. Typical results of SOT-induced magnetization reversal are shown in Fig.~\ref{Fig5} for respective positive $H_x=+250, 373, 495, 608$~mT (Fig.~\ref{Fig5}a) or equivalent negative $H_x=-240, -368, -495, -600$~mT (Fig.~\ref{Fig5}b) fields for $t_{Al}=3$~nm. Importantly, a complete magnetization reversal is found for $J_c\approx \pm 2-2.5 \times 10^{11} A/m^2$ current densities. We notice that the polarity of magnetization reversal cycle depends on the sign of $H_x$ as expected from the SOT symmetry. We also find that the critical current density decreases as $|H_x|$ increases as also observed in Refs.~\cite{rojas2016,figuereido2021} with typical current density reaching $J_c=3-3.5 \times 10^{11}A/m^2$. In Fig.~\ref{Fig5}c, we display a comparison of magnetization reversal cycles for different Al thickness $t_{Al}=1, 2, 3$ and 4~nm, acquired at $H_x=+250$~mT. It is to be noticed that the devices having $t_{Al}= 3$ and 4~nm show a complete switching of magnetization (between P and AP states) whereas the ones with $t_{Al}= 1$ and 2~nm exhibit only partial reversal, as expected from SOT measurements shown in Fig.~\ref{Fig3}. We can hence conclude that increasing the Al thickness in Pt(8)|Co(0.9)|Al($t_{Al}$)|Pt(3) series increases the SOT allowing a reduction of the critical current density for complete magnetization switching by more than 30\%.

\vspace{0.1in}


From fundamentals, SOT is generally determined \textit{via} the real and imaginary parts of the spin-mixing conductance~\cite{brataas2006,amin2016,amin2016b,cosset2021} for thick magnetic layers. For $t_{\rm Co}$ smaller than the decoherence length, the description however becomes much more complex~\cite{kim2019}. In order to tackle this issue that corresponds precisely to our experimental conditions, we have developed a model and numerical analyses (\textcolor{blue}{(SI-VII)}) gathering most of the ingredients discussed above. To this aim, we have adapted the generalized drift-precession-diffusion equations~\cite{levy2003,waintal2012,haney2013,kim2019} in agreement with the spin-dependent Boltzmann theory~\cite{amin2016,amin2016b}. We provide here a subsequent insights for $H_{\text{DL}}$ and $H_{\text{FL}}$, considering the integrated value of the torque $\mathbf{\tau}_{SOT}$ by \textit{s-d} exchange ($J_{xc}$). By reciprocity, the SOT acting on the Co layer by the out-of equilibrium spin $\hat{\mathbf{\mu}}_F$ in Co writes~\cite{nikolic2018}:

\begin{equation}
    \mathbf{\tau}_{\rm SOT}=\int_\mathcal{V}\frac{d \hat{\mathcal{M}}}{d t}d\mathcal{V}\approx \int_\mathcal{V}\frac{\hat{\mathbf{\mu}}_F\times\hat{\mathbf{m}}}{\tau_J}d\mathcal{V}.
\end{equation}
where $\mathcal{V}$ is the volume of Co and $\mu_F$ is expressed in the same unit ($\mu_B/\mathcal{V}$ ) than $\hat{\mathcal{M}}$ and $\tau_J=\frac{\hbar}{J_{xc}}$. Based on our present knowledge for Co|Pt interface~\cite{jaffers2014,jaffres2020,jaffres2021}, the predicted SOT fields are in excellent agreement with the experimental ones for both DL and FL (see dashed lines in Fig.~\ref{Fig2} and  Fig.~\ref{Fig3}). We have then extracted the precession length $l_J=v_F\tau_J=0.75$~nm and a transverse decoherence length $l_\perp=v_F\tau_\Delta=1.7$~nm in the range of the values given in Ref.~\cite{waintal2012} for Co with $\mathcal{T}_{\rm Co|Pt}=\frac{G_{s}^{\rm Co|Pt}}{G_{sh}}=0.8$. Here, $G_s$ is the spin surface conductance and $G_{sh}$ is the Sharvin conductance. The spin resistances were previously estimated at $\tilde{r}_s^{\rm Pt}=2$ and $\tilde{r}_s^{\rm Al|Pt}=3$ in unit of $\frac{1}{G_{sh}}$ (\textcolor{blue}{SI-VII}). We thus point out a clear difference between Co|Al and Co|Cu in terms of the transmission coefficient at the top Co interface with Al or Cu: $\mathcal{T}_{\rm Co|Cu|Pt}=\frac{G_{s}^{\rm Co|Cu|Pt}}{G_{sh}}=0.5$ and $\mathcal{T}_{\rm Co|Al|Pt}=\frac{G_{s}^{\rm Co|Al|Pt}}{G_{sh}}=0.2$. As aforementioned, such large $\mathcal{T}_{\rm Co|Cu|Pt}$ was also deduced from AHE from a previous theory~\cite{dang2020}.

To fit the experimental data, we account for a spin-accumulation $\mu_F=\mu_{REE}$ at Co|Al(Cu) interface due to REE. For Pt|Co|Cu|Pt, we consider that $\mu_{REE}$ is \textit{homogeneous} in Co and constant for all $t_{\rm Co}$. We quantify that REE contributes $\approx$~45\% to the FLT ($\approx$~55\% to SHE) for $t_{\rm Co}$=~0.4~nm in the reference Cu samples together with a respective Larmor precession length $\lambda_J$=4.7~nm and decoherence length $\lambda_\perp=1.7$~nm. $\lambda_J$ is then found in very close agreement with a recent literature for NiFe from orbital Hall effects measurements~\cite{Ando2022}. For the Pt|Co|Al|Pt series, a much larger spin accumulation $\mu_{REE}$ is required to fit the data with a certain dependence on $t_{\rm Co}$ varying within a length scale of the order of 0.75~nm. Consequently, the FLT contribution from REE reaches $\approx$~75\% for $t_{\rm Co}=0.55-1.2$~nm ($\approx$~25\% from SHE), whereas it contributes only by 60~\% for $t_{\rm Co}$=1.4~nm ($\approx$~40~\% from SHE). Note that the large FLT in Pt|Co|Al|Pt may originate from the significant work function difference between Co and Al associated to the subsequent charge transfer giving rise to a strong interfacial electric field. In fact, we recently referred to such effect in order to elude a large Dzyaloshinskii-Moriya interaction (DMI) found in Pt|Co|Al~\cite{ajejas2021}. 
\vspace{0.1in}


In conclusion, we observe that, in atomically thin Co layers, SOTs are indeed only partially controlled by spin-current generation via the \textit{usual} SHE mechanism from a neighboring Pt layer. As a matter of fact, we find is that the amplitude of SOTs, both for Damping- and Field-like torques, varies significantly in case a light element, that is either Al or Cu, is deposited on top of this Co film, surpassing the values existing in literature. One of the most striking results, found specifically in Pt|Co|Al system, is a giant increase in the field-like component and the corresponding ratio of $H_{\text{FL}}/H_{DL}$ at Co|Al|Pt interfaces upon reducing Co thickness within the single atomic thickness range. Combined with the analytical calculations, this observation demonstrates that Rashba interaction exists, and even dominates the bulk SHE contribution, at all-metallic Co|Al interface, a mechanism of spin-current generation that is rather known at oxide interfaces, in 2DEG systems and at surface states of topological insulators. Our experimental findings suggest a significant role of charge transfer phenomena between interfaces as well as the presence of large interfacial spin-orbit related conversion effects at Co|Al interface. Beyond the fundamental interest, the present demonstration of its impact in all-metallic systems in the limit of atomically thin ferromagnets opens also a new research direction for CMOS-compatible and cost-effective spinorbitronic based technology.

\section*{Acknowledgments}

This work has been supported by DARPA TEE program grant (MIPRHR - 0011831554), ANR grant TOPSKY (ANR-17-CE24-0025), the FLAG - ERA SographMEM (ANR-15-GRFL-0005) and the Horizon2020 Framework Program of the European Commission, under FETProactive Grant agreement No. 824123 (SKYTOP) (H2020 FET proactive 824123).


\providecommand{\noopsort}[1]{}\providecommand{\singleletter}[1]{#1}%

\end{document}